\documentclass[prb, aps, twocolumn,showpacs,amsmath,amssymb, superscriptaddress, 10pt]{revtex4-2}

\usepackage{amsmath}    
\usepackage{amssymb}
\usepackage{graphicx}   
\usepackage{xcolor}      
\usepackage[normalem]{ulem}
\usepackage{bm}
\usepackage[version=3]{mhchem}
\usepackage{hyperref}   

\hypersetup{colorlinks,linkcolor=blue,urlcolor=blue,citecolor=blue}

\newcommand{\avg}[1]{\langle #1\rangle}

\usepackage{ulem}

\begin{document}
\author{Bogdan Ostahie}
\affiliation{National Institute of Materials Physics, 077125 Bucharest-Magurele, Romania}
\author{Doru Sticlet}
\email{doru.sticlet@itim-cj.ro}
\affiliation{National Institute for R\&D of Isotopic and Molecular Technologies, 67-103 Donat, 400293 Cluj-Napoca, Romania}

\author{C\u{a}t\u{a}lin Pa\c{s}cu Moca}
\affiliation{Department of Theoretical Physics, Institute of Physics, Budapest University of Technology and Economics, 	M\H{u}egyetem rkp.~3, H-1111 Budapest, Hungary}
\affiliation{Department  of  Physics,  University  of  Oradea,  410087,  Oradea, Romania}

\author{Bal\'azs D\'ora}
\affiliation{Department of Theoretical Physics, Institute of Physics, Budapest University of Technology and Economics, M\H{u}egyetem rkp.~3, H-1111 Budapest, Hungary}
\affiliation{MTA-BME Lend\"ulet Topology and Correlation Research Group, M\H{u}egyetem rkp.~3, H-1111 Budapest, Hungary}

\author{Mikl\'os Antal Werner}
\affiliation{Department of Theoretical Physics, Institute of Physics, Budapest University of Technology and Economics, M\H{u}egyetem rkp.~3, H-1111 Budapest, Hungary}
\affiliation{MTA-BME Quantum Dynamics and Correlations Research Group, M\H{u}egyetem rkp.~3, H-1111 Budapest, Hungary}

\author{J\'anos K. Asb\'oth}
\affiliation{Department of Theoretical Physics, Institute of Physics, Budapest University of Technology and Economics, M\H{u}egyetem rkp.~3, H-1111 Budapest, Hungary}
\affiliation{Institute for Solid State Physics and Optics, Wigner Research Centre for Physics, P.O. Box 49, H-1525 Budapest, Hungary}

\author{Gergely Zar\'and}	
\affiliation{Department of Theoretical Physics, Institute of Physics, Budapest University of Technology and Economics, M\H{u}egyetem rkp.~3, H-1111 Budapest, Hungary}
\affiliation{MTA-BME Quantum Dynamics and Correlations Research Group, M\H{u}egyetem rkp.~3, H-1111 Budapest, Hungary}

\title{Multiparticle quantum walk: A dynamical probe of topological many-body excitations}

\begin{abstract}
Recent experiments demonstrated that single-particle quantum walks can
reveal the topological properties of single-particle states. Here, we
generalize this picture to the many-body realm by focusing on
multiparticle quantum walks of strongly interacting fermions.  After
injecting $N$ particles with multiple flavors in the interacting
SU$(N)$ Su-Schrieffer-Heeger chain, their multiparticle continuous-time quantum walk is monitored by a variety of methods. We find that
the many-body Berry phase in the $N$-body part of the spectrum signals
a topological transition upon varying the dimerization, similarly to
the single-particle case. This topological transition is captured by the single- and many-body mean chiral displacement during the quantum walk and remains present for strong interaction as well as for moderate disorder.
Our predictions are well within experimental reach for cold atomic gases and can be
used to detect the topological properties of many-body excitations
through dynamical probes.
\end{abstract}
\maketitle

\section{Introduction}
Our conventional understanding of phase transitions associated with a symmetry breaking and the emergence of a local order parameter~\cite{Landau1980} has been extended with the advent of topological insulators~\cite{Hasan2010, Qi2011}.
Such materials are \emph{not} characterized by a local order parameter and display various topological phases identified by topological invariants~\cite{Kitaev2009, Chiu2016}.
In noninteracting systems, topological invariants account for the topological character of single-particle wave functions and are accompanied by
robust low-energy features at the boundaries of the system~\cite{Essin2011,Prodan2018}.
Not only are these useful in revealing the topological properties of matter, but also hold the promise 
to revolutionize quantum computation, quantum technologies, and spintronics~\cite{Alicea.2011,Smejkal.2018,He.2022}.

While the study of noninteracting topological systems has advanced significantly in recent years, 
and the basic physics of noninteracting topological insulators is well understood by now, 
research into the analogous strongly correlated systems has progressed slowly~\cite{Wang2014,Rachel2018}, and it is rather unclear whether topology survives the presence of strong interactions~\cite{Gurarie2011}.
Electron-electron interactions may in some situations be responsible for topological phase transitions,  
as it was demonstrated that a single quadratic band crossing is unstable with respect to topological insulating phases in the presence of interactions~\cite{kaisun, kunyang}.  
More generally, a mean-field decoupling of the interaction can result in an effective spin-orbit coupling, for instance, thus inducing a transition from a topologically trivial to a
nontrivial phase~\cite{raghu, vozmediano,igorPRB}.  
Thus, understanding the topology of the ensuing phase and disentangling it from
single-particle topological states is far from trivial. 

It is therefore extremely important to establish ways for 
characterizing an interacting topological insulator~\cite{Ludwig2015, Rachel2018}.
The most likely candidate, which generalizes the underlying noninteracting Berry phase, would be the occurrence of a quantized many-body Berry phase~\cite{Berry1984, Resta2000, Xiao2010, Vanderbilt2018}.
Other proposals suggest a connection between topology and the degeneracy of the entanglement ~\cite{Pollmann2010,Fromholz2020} or entropy itself~\cite{Wang2015}. However, 
testing any of these hypotheses experimentally in bulk systems is a challenging task.

A simple, experimentally appealing  way  to probe topology invokes the quench dynamics of quantum particles in single-particle quantum walks~\cite{Cardano2017}.
Recent experiments have revealed the presence of bound states at the interface of systems with
different topological phases~\cite{Kitagawa2012,Groh2016} and allow for
the detection of topological invariants in cold atoms~\cite{Atala2013}
or in nanophotonic topological lattice through the
mean chiral displacement (MCD)~\cite{Cardano2017,Meier2018,Wang2019}, 
which measures the difference of the average occupations 
of the dimerized lattice in the long-time limit [see Eq.~\eqref{eq:P1} for a concise definition].
So far, the MCD has been measured experimentally only in a noninteracting setup, 
by performing single-photon quantum walks in topological photonic lattices~\cite{Cardano2017, Rakovszky2017, Zhan2017, Wang2019}. 

By now, experimental quantum walks have been implemented for trapped atoms and ions~\cite{Schmitz2009,Karski2009}, photons~\cite{Peruzzo2010, Schreiber2010,Poulios2014}, or spin impurities~\cite{Fukuhara2013, Fukuhara2013a}, and full control over the dynamics has been achieved. 
Furthermore, with  recent advances in nanophotonics~\cite{Lu2014, Ozawa2019}, 
quantum walks of correlated photons, and correlation effects in Bloch oscillations~\cite{Preiss2015, Geiger2018} in multiparticle quantum walks have already been realized. 
It is therefore interesting to explore whether the MCD
is a suitable quantity to capture  topology in the presence of strong interactions 
as well~\cite{Haller2020}.

\section{Models}
\subsection{Interacting \texorpdfstring{SU($N$)}{SU(N)} Su-Schrieffer-Heeger model}
In the present paper we address this problem, and investigate the effect of 
interactions on the bulk states' topology in a multiparticle quantum walk setup.
For that, we corroborate  results on the Berry phase, computed using the 
many-body spectrum in a subspace of excited states with a reduced number of particles, 
and results on single-particle and many-body MCDs. 

\begin{figure}[t]
  \centering
  \includegraphics[width=0.8\columnwidth]{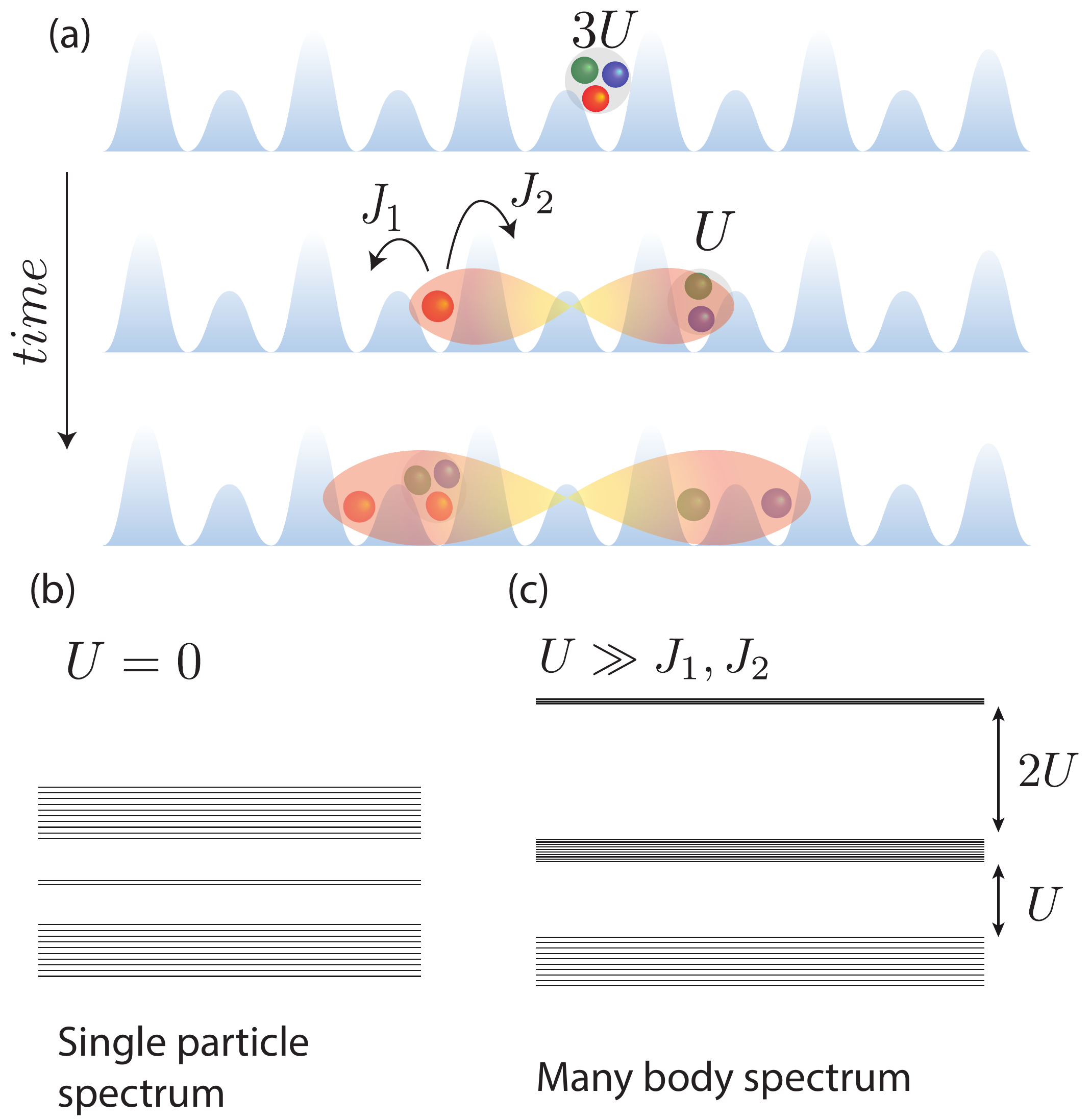}
  \caption{ 
  (a) Multiparticle quantum walk  on an SU(3) SSH lattice. 
  $U$ represents the strength of the on-site interaction. 
  Three particles are injected at $t=0$.
  The single- or multiparticle MCDs measured at later times provide information on
  the bulk states' topology. Here, $J_{1,2}$ represent the hoppings of the dimerized lattice, while $U$ 
  is the on-site repulsion energy.
  (b) Typical single-particle spectrum in the topological regime 
  of the noninteracting spinless SSH model, with mid-gap lines indicating topological edge states. 
  (c) Three-particle many-body spectrum of the SU(3) model for $U\gg J$. The spectrum consists 
  of three bands separated by energy gaps of  order  $\sim U$.}
  \label{fig:sketch}
\end{figure}

For the noninteracting SU$(N)$ Su-Schrieffer-Heeger (SSH) model, the cases of $N$  odd and even differ significantly, similarly to other spin models~\cite{haldane}.
The Berry phase $\gamma_B$ is marked by a jump of $\pi N$ (modulo $2\pi$) at the topological transition, 
therefore for even $N$, $\gamma_B$ is unable to distinguish between the topological 
phases~\cite{Verresen2017}.  
This is inherited in the interacting version of the model, 
where the many-body Berry phase defined below displays a jump $\pi$ only for odd $N$, and can be used as an indicator of the topological transition.
Nevertheless, the dynamically accessible single- and many-body MCDs identify correctly the topological  phases for any number of flavors~\cite{SM}, are robust against moderate disorder, and serve as
suitable measures to interrogate the bulk states' topology.
For clarity, in this paper we mostly explore the simplest 
case, $N=3$, where $\gamma_B$ is also a good indicator.
Results for the SU(2) SSH model are detailed in the Supplemental Material (SM)~\cite{SM}, where  
extensions to $N>3$ are also considered.

The prototypical  SU($N$) SSH model~\cite{Su1979} with an on-site Hubbard interaction is given by
\begin{eqnarray}
    H &=& \sum_{x=-L/2}^{L/2} \bigg\{-J\sum_{\alpha=1}^{N}  [1+(-1)^x \delta](c_{x,\alpha}^\dagger c_{x+1,\alpha}^{} +\text{h.c.})\nonumber\\
    && {}
    + U  \sum_{1\le\alpha<\beta \le N} c_{x,\alpha}^\dag c_{x,\alpha}^{}c_{x,\beta}^\dagger c_{x,\beta}^{}
    \bigg\}\, .
    \label{eq:H_SSH}
\end{eqnarray}
The first term in Eq.~\eqref{eq:H_SSH} accounts for the dimerized hopping with amplitudes $J_{1,2}=J(1\pm \delta)$ between neighboring sites (see Fig.~\ref{fig:sketch}),  with
$c^{(\dagger)}_{x,\alpha}$  the annihilation (creation) 
operator of a fermion with flavor $\alpha$ at site $x$. 
The second, the Hubbard term, describes the on-site interaction between fermions with different flavors. 

Interacting extensions of the SSH model have been used so far as a springboard 
to understand  the effects of strong interactions, and have been concerned either with 
interacting bosonic~\cite{DiLiberto2016, Gorlach2017, Stepanenko2020}, spinless fermionic models~\cite{Sirker2014}, or  SU(2) fermionic models,i.e., with flavors associated with the $\uparrow(\downarrow)$ spin labels~\cite{Manmana2012,Sirker2014,Wang2015,Le2020}. 

In the absence of interactions, $U=0$, the different flavor channels
are decoupled and for $N=3$ the model reduces to three copies of the
noninteracting spinless SSH model~\cite{Su1979}.  
The noninteracting model with $U=0$ has an antiunitary chiral symmetry $\Gamma$, which transforms the local operators as $\Gamma c_{x, \alpha}\,\Gamma^{-1}\to (-1)^{x}
c_{x, \alpha}^{\dagger}$ and $\Gamma\, i\, \Gamma^{-1} \to -i $
(see SM for details), and 
the model displays a topological phase transition at $\delta=0$,
from a trivial ($\delta<0$)  to a topologically nontrivial ($\delta>0$) phase. 
The transition is characterized by a jump of $\pi$ in the Zak phase~\cite{Atala2013}.
A typical band structure for the spinless SSH model in the topological regime, with the zero-energy  modes emphasized, is displayed in Fig.~\ref{fig:sketch}(b).
The interacting Hamiltonian~\eqref{eq:H_SSH} also respects chiral symmetry apart from an overall chemical potential term,  which is, however, irrelevant for the dynamics in a closed system, investigated here.

In the  interacting model  we find that  excitations 
still exhibit two  topological phases due to the presence of inversion symmetry, 
although  zero-energy topologically protected edge excitations cease to exist.
This is similar to the situation in noninteracting systems, where inversion symmetry 
enforces the Zak phase quantization, $\gamma_Z=0$ or $\pi$ (modulo $2\pi$)~\cite{Zak1989}. 
This quantization has recently been demonstrated experimentally in a photonic lattice, where the chiral symmetry of a noninteracting SSH model is broken by engineered long-range hopping, designed to preserve inversion symmetry~\cite{Jiao2021}.

To explore the topology of excitations, we inject three particles into the empty lattice 
in the middle of the chain  by the three-particle creation operator 
$\Phi_{x}^{(3)\dagger} =\prod_{\alpha=1}^3 c^\dagger_{x, \alpha}$,  and  investigate
 the effect of interactions in the quench dynamics in a multiparticle quantum
  walk setup~\cite{Aharonov1993,Farhi1998,Childs2002}, 
  where the wave function follows a unitary evolution 
  $| \Psi(t)\rangle =e^{-iHt} |\Psi(t=0)\rangle$  [see Fig.~\ref{fig:sketch}(a)]. 

\subsection{Effective model for trions} 
Before investigating the quench dynamics, let us discuss shortly the many-body ``band structure'' of the Hamiltonian~\eqref{eq:H_SSH}.
A sketch of this band structure is displayed for $N=3$ in Fig.~\ref{fig:sketch}(c)  in the limit of strong interactions, $U\gg J$. 
The lowest band is constructed from  states of propagating single fermions,
and  has a width   $W_1\approx 4 J$.  
The highest in energy band, the trionic band, is constructed  from  
states in which all the three particles  reside mostly at the same site~\footnote{Trion states are defined as three-particle eigenstates of the interacting  Hamiltonian, adiabatically connected to the three-particle states in the $J\to0$ limit. Doublons are defined similarly, as two-particle eigenstates.}. 
This band consists of $L$ states (split into two subbands, as we show below), with very large energies
$\approx 3U$, and a small bandwidth, of the order $W_3\approx  J^3/U^2$.
Between the single-particle and the trionic bands there is the doublonic band, well separated in energy from the others. 
The average energy of doublons is of the order $\sim U$
(see Ref.~\cite{werner2022spectroscopic} and \cite{SM} for more details on the band structure). 

The initial state $|\Psi(0)\rangle = \Phi_{x}^{(3)\dagger}|0\rangle$ 
has an energy $\approx 3 U$, and has most of its weight in the trionic band. 
In the large $U$ limit, trions propagate across the lattice by high-order tunneling processes, 
and their dynamics is described by the effective Hamiltonian,
\begin{eqnarray}
    H_3 &=& \sum_{x} \Big \{  J_3[1+\delta_3(-1)^x]
    (\tilde \Phi_{x}^{(3)\dagger} \tilde \Phi_{x+1}^{(3)}+\text{h.c.})\nonumber \\  
    &&{}+ E_3  \tilde \Phi_{x}^{(3)\dagger}  \tilde \Phi_{x}^{(3)} \Big \},
    \label{eq:Heff}
\end{eqnarray}
where the $\tilde \Phi^{(3)\dagger}$ create dressed trion states, and  $E_3$ and $J_3$ are the effective on-site energy and hopping, describing the trionic band~\footnote{The operators $\tilde \Phi_{x}^{(3)\dagger}$ and $\Phi_{x}^{(3)\dagger}$ should not be confused, as the first one creates a dressed trion state while the second one is a three-particle creation operator at position $x$}. 
The effective Hamiltonian \eqref{eq:Heff} preserves the bipartite nature of the original 
Hamiltonian~\eqref{eq:H_SSH} with an effective dimerization $\delta_3$ of the same order of magnitude as $\delta$~\cite{SM}. 
Trions are extremely heavy objects, and propagate through the lattice with a small effective velocity $v_3\approx 2\,J_3\ll v_1$, much smaller than the single-particle speed of propagation, $v_1=2J(1-|\delta|)$.
The initial three-particle state $\Phi_{x}^{(3)\dagger}|0\rangle$ overlaps with a large probability
$p_3 \approx  1$ with the dressed trionic state,  $\tilde\Phi_{x}^{(3)\dagger}|0\rangle$.
It contains, however, with a small probability $p_2\approx \frac32 J^2/U^2 $ an admixture of doublons and free fermions, and an even smaller contribution from three independent fermions. 
Although they have a small contribution, these latter components propagate fast compared to trions.

\section{Topological transitions in the interacting \texorpdfstring{SU(3)}{SU3} SSH model}
\subsection{Many-body Berry phase}
A numerical analysis performed by computing the winding number using the Green's functions for the SU($N=2$) version of the model~\eqref{eq:H_SSH} at half filling indicates that interactions do not destroy the topology, although the bulk--boundary correspondence no longer survives~\cite{Manmana2012}.
Similar conclusions have been drawn from the analysis of two-body physics in spinless bosonic SSH models~\cite{DiLiberto2016,Gorlach2017}.

We now investigate the bulk properties of SU(3) SSH systems using the many-body Berry phase $\gamma_B$. 
We compute $\gamma_B$ by using an approach, 
where in contrast to the standard methods~\cite{Resta1995,Ortiz1996,Resta1999,Resta2000}, we determine the Berry phase over a \emph{subset} of excited many-body states instead of simply using the ground state.
The justification to use such a subset is that the initial three-particle  state is built out of highly excited eigenstates. 
We consider a ring geometry, and impose twisted boundary conditions on the many-body spectrum.
This is done by modifying one of the hopping terms $J\to Je^{i
\theta_n}$ with  $\theta_n=2\pi n/M$, $n\in\{0,1,\dots,M-1\}$, and $M$ controlling the twist angle discretization.
For each $\theta_n$, we diagonalize the Hamiltonian~\eqref{eq:H_SSH} within the $N=3$ subspace, 
$H(\theta_n)\Psi^{(n)}_j = E_j^{(n)}\Psi^{(n)}_j, $
and obtain the full many-body spectrum $\{E_j^{(n)}\}$ of the Hamiltonian.

The many-body Berry phase is well defined over a subset $\{\Psi_j\}$ of the many-body spectrum,
 separated by a gap from the rest of the states for all twist angles. 
 By generalizing the procedure used  for the ground 
 state~\cite{Resta1995, Resta1998, Resta1999, Aligia1999, Resta2000}, we obtain
\begin{equation}
\gamma_B =-\text{Im}\log\prod_{n=0}^{M-1} \det [S^{(n,n+1)}],
\end{equation}
where the elements of the matrix $S^{(n,n+1)}$  are
\begin{equation}
S^{(n,n+1)}_{jj'} = \avg{\Psi_j^{(n)}|e^{2\pi i X/ML}|\Psi_{j'}^{(n+1)}},
\end{equation}
with $j,j'$ indexing the subset of many-body states, $\{\Psi_j\}$, and  with  $X = \sum_{j,\alpha} x \,c^\dagger_{x,\alpha} c_{x,\alpha} $ 
the many-body position operator along the chain.
Under inversion symmetry the Berry phase transforms as $\gamma_B\to-\gamma_B$ (modulo $2\pi$). Therefore, even in the interacting system,  inversion symmetry enforces $\gamma_B=0$ or $\pi$.

\begin{figure}[t]
	\includegraphics[width=0.8\columnwidth]{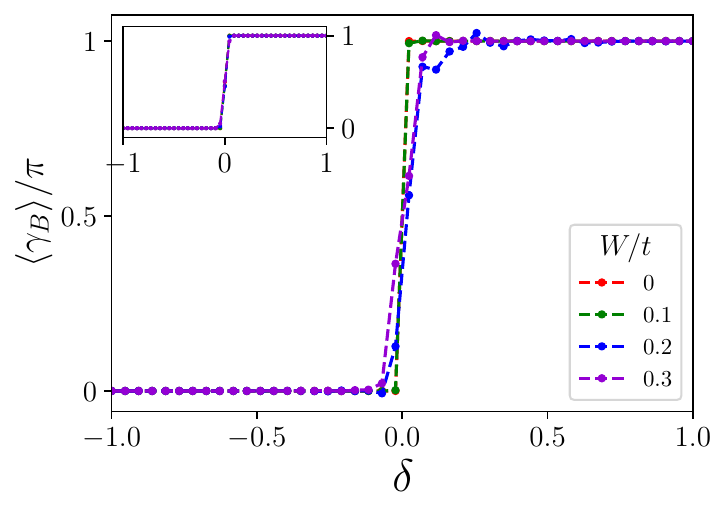}
	\caption{  Berry phase $\gamma_B$ as a function of $\delta$, 
		obtained by integrating half of the excited (trion) many-body states in 
		an $L=8$ lattice with twisted boundary conditions, with $M=20$ and
		$U=3J$, for different chiral disorder strengths $W$. 
		For each $W\ne0$, we average over 100 disorder realizations.
		Inset: Berry phase for the noninteracting SSH model
		with  hopping disorder.}
	\label{fig:berry_phase}
\end{figure}

As explained earlier, the initial three-particle state is mainly constructed from states within the trionic band,  which  contains $L$ states divided into two subbands, separated 
by a gap generated by the effective dimerization parameter $\delta_3$. 
By including  the $L/2$ highest-energy many-body states 
 into the set $\{\Psi_j\}$, we recover a jump of $\pi$ in $\gamma_B$ at $\delta=0$
 (see  Fig.~\ref{fig:berry_phase}).    
This indicates a topological phase transition
 between two  topologically distinct regions,  $\delta \lessgtr 0$. 

\subsection{Single- and many-body mean chiral displacements}
The mean chiral displacement  represents a dynamical measure capable to distinguish
between the topological and the trivial regimes~\cite{Cardano2017}, defined as
\begin{equation}
{\cal P}_1(t)= \sum_x  \langle \Psi(t)|  \; (x-x_0) \;\Gamma \,n(x)  |\Psi(t)\rangle.
\label{eq:P1}
\end{equation}
Here, $p(x)= (x-x_0)\,  n(x)$ denotes the regular polarization operator with 
respect to the reference point $x_0$ and by using the chiral-symmetry operator $\Gamma $, distinguishing between different sublattices,
the ``chiral polarization'' operator or MCD is $(x-x_0) \Gamma n(x)$.
$|\Psi(t)\rangle$ is the wave function time evolved from an appropriate initial state, localized on site $x_0$,  now set as  the middle of the chain~\cite{Haller2020}.
Although ${\cal P}_1(t)$ depends on time, for noninteracting systems it converges to ${\cal P}_1(t\to \infty)\simeq \nu/2$, where $\nu$ is the bulk topological invariant associated with the 
Zak phase (see SM~\cite{SM} for a derivation). 

Since the many-body state is mainly constructed from states in the trionic band,
it is  natural to extend the 
definition of the noninteracting MCD in Eq.~\eqref{eq:P1},
and introduce 
the many-body MCDs as
\begin{equation}
	{\cal P}_3(t) \equiv 
\frac 1 {\avg{n_3(t)}}
\langle \Psi(t) | 
\sum_{x} (-1)^x x\,
\Phi^{(3)\dagger}_{x} \Phi^{(3)}_{x}
|\Psi(t) \rangle\;.
\label{eq:P13}
\end{equation}

The prefactor $\langle n_{3}(t)\rangle\leq 1$ measures the probability of three-particle 
occupation,  $\langle n_{3}(t)\rangle  =\sum_x \langle \Psi(t)| \,\Phi_3^\dagger (x) \Phi_3(x) |\Psi(t)\rangle $ \footnote{In Eq.~\eqref{eq:P13} the reference point is the set to the middle of the chain and corresponds to $x_0 =0$.}. 
Simple perturbation theory in $J$ yields the asymptotic  estimate,  $\langle n_3(t\to \infty)\rangle\sim 1-{\cal O}(J^2/U^2)$.
The results for  $\gamma_B$ are corroborated by the single-particle and the many-body MCDs, ${\cal P}_{1,3}(t)$,
computed using both exact diagonalization and time evolving block decimation (TEBD)~\cite{Vidal2007}.  
To tame the time oscillations, we also investigate the cumulative average MCDs, 
${\cal P}^c_{1,3}(t)=\int_0^{t} {\cal P}_{1,3}(t') dt'/t$. 
Notice that $ {\cal P}_{1,3}(t)$ are measures of the bulk topology, and do not 
provide any information on the localized states at the edges. Boundaries produce  spurious artifacts while measuring $ {\cal P}_{1,3}(t)$,
they reflect propagating fronts, and yield boundary-induced oscillations in the simulations.
The measurement time $t_{\infty}$ is therefore always set  to be smaller in our simulations than the time required for the front to reach the system boundaries.
 
\begin{figure}[t]
    \includegraphics[width=\columnwidth]{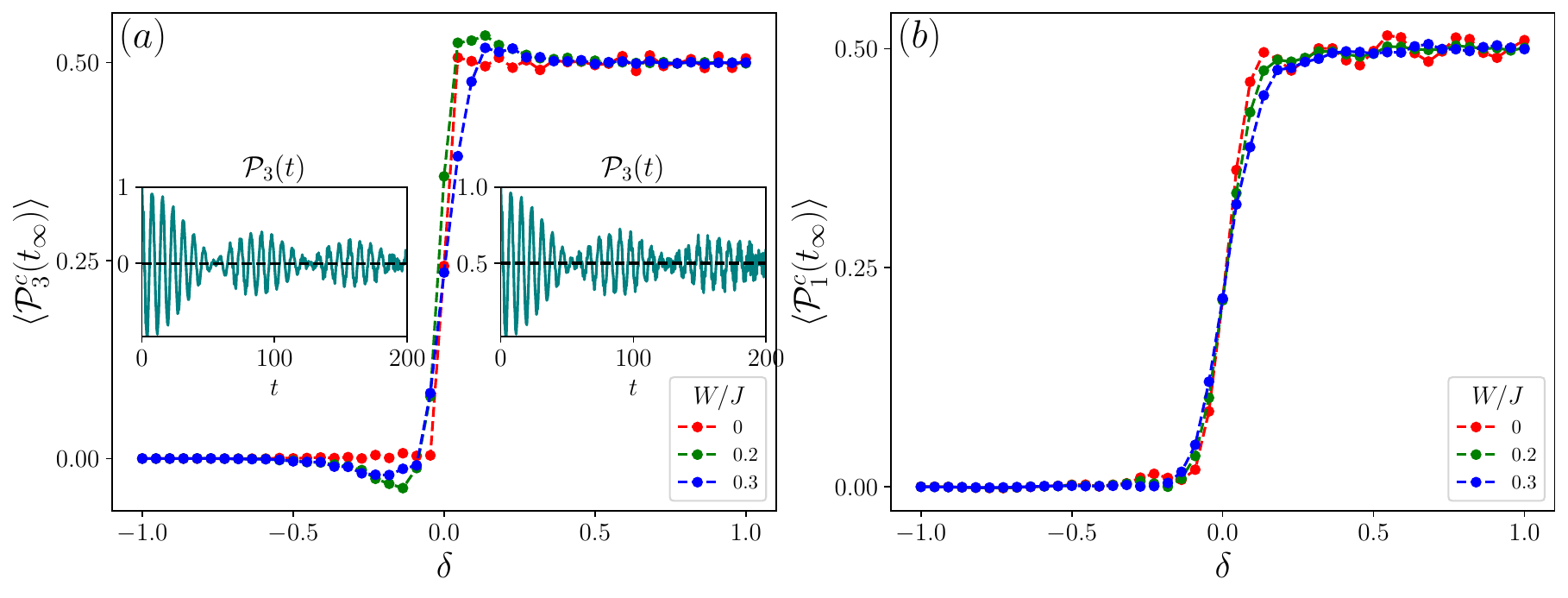}
    \caption{Mean chiral displacement in the interacting SU(3) SSH model. (a) Cumulative average MCD $\mathcal P^c_3$ for different chiral-symmetry preserving disorder strengths $W$ as a function of dimerization parameter $\delta$, for lattice size $L=30$, and $t_{\rm \infty}=40/J$.
    Insets: Typical evolution of the MCD $\mathcal P_3$ in the clean system at (right) $\delta=0.5$ and (left) $\delta=-0.5$.
    (b) Cumulative disorder-averaged  MCD  $\mathcal P^c_1$ for different disorder amplitudes as a function of dimerization parameter $\delta$, for $L=40$, and $t_{\rm \infty}=20/J$. In both panels $U=3J$, and the average is done over 100 disorder realizations at each $W$.}
    \label{fig:trions}
\end{figure}

The MCDs  $ {\cal P}_{1,3}(t)$ are presented in Fig.~\ref{fig:trions}.
The two insets in Fig.~\ref{fig:trions}(a) display the time evolution of ${\cal P}_3(t)$ in the two regimes, as computed with  TEBD.
The asymptotic limit ${\cal P}_3(t_\infty)$ depends on the value of $\delta$, 
and it converges to $0$ $(0.5)$ in the trivial (topological) regime. 
The main panel shows the $\delta$ dependence of the asymptotic values. 
The small oscillations in the asymptotic values are due to  
finite-size effects and limited simulation runtime, and they reduce in 
amplitude with increasing system size.

As clear from Fig.~\ref{fig:trions}, the many-body and the single-particle MCDs are both suitable tools to differentiate between topologically distinct regions. 
We also demonstrate~\cite{SM} that both MCDs remain reasonably well quantized even away from the strongly correlated region as well.

\subsection{Disorder effects}
The quantization of  the MCD is relatively robust against disorder. 
We have tested this robustness against different kinds of disorder.   We
have introduced a moderate \emph{chiral} (hopping) disorder
$W$ breaking the inversion symmetry to test the robustness of
$\gamma_B$.  Fig.~\ref{fig:berry_phase} displays the results for
$\gamma_B$ for different $W$'s.  Although the jump at the transition becomes somewhat smeared,  $\gamma_B$ remains quantized as long as $W\lesssim \delta_3\propto \delta$.
 
Similarly, disorder does not affect significantly either ${\cal P}_{1}(t_\infty)$ 
or ${\cal P}_{3}(t_\infty)$, and the MCDs remain  robust as well against the on-site or hopping disorder, as long as disorder is small compared to the topological gap for the excitations 
(see Fig.~\ref{fig:trions}). 
Although moderate disorder does not destroy the topological band of excitations, strong interactions reduce the trion bandwidth as $\sim J^3/U^2$, 
and topological trion excitations become more  susceptible to disorder in the limit of strong interactions (see SM~\cite{SM}).

\subsection{Extension to the \texorpdfstring{SU($N$)}{SU(N)} SSH model.} 
For clarity, so far we mostly focused on  the SU(3)  SSH model.  
Many of the results carry over, however, to the general SU($N$) case.
In the latter case, the $N$-particle many-body spectrum has a ladderlike structure with $N$ bands, separated by energy gaps of the order of $U$. 
The highest in energy band, the $N$-ion band, has an energy $E_N\approx N(N-1)U/2$.
The construction of the effective model that describes the $N$-ionic band, similar to  Eq.~\eqref{eq:Heff} for the trions, is detailed in Refs.~\cite{SM, werner2022spectroscopic}. 
Computing the many-body Berry phase and the polarization is done along the same lines as for SU(3).
Introducing the $N$-particle creation operator, $\Phi_x^{(N)} = \prod_{\alpha=1}^N c^{\dagger}_{x,\alpha}$, the many-body MCD is obtained as in Eq.~\eqref{eq:P13} with the substitution $3\to N$. Details for the Berry phase and MCD in the SU(2) case are presented in Ref.~\cite{SM}. 
Our findings also  carry over to the case when 
\emph{several} trions are injected to the lattice. 
We performed calculations for the case where two trions have been injected, and our findings~\cite{SM} indicate that, irrespective of  the strong interaction between trions, the MCDs 
${\cal P}_{1,3}(t_\infty)$ remain unaffected in the long-time limit, and capture correctly the topological transition.
\subsection{Connection to experiments}
Experimental implementation of repulsive bound pairs has been realized
by loading \ce{^{87}Rb} atoms in an optical lattice~\cite{Winkler.2006},
while later on, the same setup was used to probe the dynamics and 
equilibrium properties of the topological SSH model~\cite{Meier.2016}.
In a state-of-the-art experiment~\cite{Hofrichter.2016},  
the SU$(N>2)$ Fermi-Hubbard (FH) model has been realized
by using \ce{^{173}Yb} atoms. By adjusting the lattice depth,
one can tune the parameters of the FH model in the range  $U/J\simeq 0.025- 2.5$,
confirming that our simulations are well within  experimental reach.
Experimentally, the three-body losses may become 
an important factor. The ratio of the on-site interaction and the three-body loss rate $\gamma$ is found to scale as $U/\gamma h \propto (\lambda/a_s)^3$ for $N=3$, where the  scattering length for \ce{^{173}Yb}, 
$a_s\approx 10\,\rm nm $  is much shorter than the wavelength $\lambda = 759\,\rm nm$ of the confining laser~\cite{Hofrichter.2016}. In this system, for an interaction strength  $U/J = 1$, we obtain the estimate
$U/h \approx  J/h \approx 300 \mathrm{Hz}$, while the three-body loss rate is only $\gamma \approx 0.16 \mathrm{Hz} \ll J/h$. 
The timescale of the three-body losses is  at least one order or 
magnitude larger than the measurement time $t_\infty$ of the experiment.
The three-particle occupation defined in
Eq.~\eqref{eq:P13} can be measured by means of quantum gas microscopy, 
by extending the method of measuring two-particle occupations\cite{hartke,mitra}. 
This is, however, not necessary in the strongly interacting regime, where $\langle n_3(t)\rangle$ is almost unity, as discussed above.

\section{Conclusions}
We have investigated the quench dynamics in a multiparticle continuous time quantum walk, and evaluated the effect of strong interactions on the topological properties of bulk excited states in the SU(3) SSH model.
Strong interactions generically violate the conventional chiral symmetry of SSH models, nevertheless, a quantized many-body Berry phase is observed in the presence of inversion symmetry, signaling two distinct topological phases, separated by a $\pi$ jump.
The many-body topological phases are robust against moderate disorder.
A similar discontinuity shows up in the asymptotic value of single-particle and  many-body mean chiral displacements.
Measuring the latter quantities in multiparticle quantum walk setups is within 
experimental reach, and could be used to infer the topological properties of the
excited states.
Similar features are expected for general SU($N$) models, as well.

\begin{acknowledgments}
This research is supported by UEFISCDI, 
under Projects No.~PN-III-P4-ID-PCE-2020-0277, under the project for funding 
the excellence, Contract No.~29 PFE/30.12.2021, 
No.~PN-III-P1-1.1-PD-2019-0595, and No.~PN-III-P1-1.1-TE-2019-0423, 
by the Core Program of the National Institute of Materials Physics, granted by the Romanian MCID under the Project No.~PC2-PN23080202,
and by the National Research, Development and Innovation Office NKFIH within the Quantum 
Technology National Excellence Program (Project No.~2017-1.2.1-NKP-2017-00001), 
K134437, K142179,  SNN139581, and   the BME-Nanotechnology FIKP grant (BME FIKP-NAT). 
M.A.W.~has also been supported by the Hungarian Academy of Sciences through the J\'anos Bolyai 
Research Scholarship and by the ÚNKP-22-5-BME-330 New National Excellence Program of the Ministry for 
Culture and Innovation from the source of the National Research, Development and Innovation Fund.
\end{acknowledgments}

\bibliographystyle{apsrev4-2}
\bibliography{bibl}
\end{document}


\author{Bogdan Ostahie}
\affiliation{National Institute of Materials Physics, 077125 Bucharest-Magurele, Romania}
\author{Doru Sticlet}
\email{doru.sticlet@itim-cj.ro}
\affiliation{National Institute for R\&D of Isotopic and Molecular Technologies, 67-103 Donat, 400293 Cluj-Napoca, Romania}

\author{C\u{a}t\u{a}lin Pa\c{s}cu Moca}
\affiliation{Department of Theoretical Physics, Institute of Physics, Budapest University of Technology and Economics, 	M\H{u}egyetem rkp.~3, H-1111 Budapest, Hungary}
\affiliation{Department  of  Physics,  University  of  Oradea,  410087,  Oradea,  Romania}

\author{Bal\'azs D\'ora}
\affiliation{Department of Theoretical Physics, Institute of Physics, Budapest University of Technology and Economics, M\H{u}egyetem rkp.~3, H-1111 Budapest, Hungary}
\affiliation{MTA-BME Lend\"ulet Topology and Correlation Research Group, M\H{u}egyetem rkp.~3, H-1111 Budapest, Hungary}

\author{Mikl\'os Antal Werner}
\affiliation{Department of Theoretical Physics, Institute of Physics, Budapest University of Technology and Economics, M\H{u}egyetem rkp.~3, H-1111 Budapest, Hungary}
\affiliation{MTA-BME Quantum Dynamics and Correlations Research Group, M\H{u}egyetem rkp.~3, H-1111 Budapest, Hungary}

\author{J\'anos K. Asb\'oth}
\affiliation{Department of Theoretical Physics, Institute of Physics, Budapest University of Technology and Economics, M\H{u}egyetem rkp.~3, H-1111 Budapest, Hungary}
\affiliation{Institute for Solid State Physics and Optics, Wigner Research Centre for Physics,  P.O. Box 49, H-1525 Budapest, Hungary}

\author{Gergely Zar\'and}
\affiliation{Department of Theoretical Physics, Institute of Physics, Budapest University of Technology and Economics, M\H{u}egyetem rkp.~3, H-1111 Budapest, Hungary}
\affiliation{MTA-BME Quantum Dynamics and Correlations Research Group, M\H{u}egyetem rkp.~3, H-1111 Budapest, Hungary}

\title{Supplemental Material for ``Multiparticle quantum walk: A dynamical probe of topological many-body excitations''}

\begin{abstract}
In this supplemental material, we detail several of the points in the main text. We derive effective models for SU(2) and SU(3) Hamiltonians in the highest excited states, in the limit of strong interactions. We discuss the role of chiral and inversion symmetries in the Berry phase quantization.
We show further numerical probes for the mean chiral displacement (MCD) in SU(2) and SU(3) models, and illustrate the time evolution of a bound two-particle or three-particle state in SU(2) and SU(3) models, respectively.
Additionally, we show that the MCD is also quantized in the case where two trions are injected in the lattice.
Finally, there is a discussion of the many-body Berry phases in SU($N$) models, when $N$ is odd or even, and when varying the subset of many-body states over which it is computed.
\end{abstract}

\maketitle

\tableofcontents

\section{Effective models}
We discuss the general situation when the lattice is described by the SU($N$) Su-Schrieffer-Heeger-Hubbard (SSHH) Hamiltonian and contains exactly $N$ particles.
We develop the effective Hamiltonian for the highest excited band with states constructed from $N$ particles with $N$ flavors (a dressed $N$-ion or just $N$-ion in short), localized on a single site, and moving in a one-dimensional lattice in the presence of Hubbard interactions.
Special attention is given to the construction of $SU(N=2)$ doublonic and $SU(N=3)$ trionic effective models. The models constructed in this way are mapped to a non-interacting SSH model with some effective couplings. The general SU($N$) SSHH Hamiltonian reads
\begin{eqnarray}\label{eq:SSHH}
H = H_U+H_T, \quad 
H_U = U \sum_{x=-L/2}^{L/2-1}\sum_{\alpha<\beta} n_{x,\alpha}n_{x,\beta},\quad 
H_T=\sum_{x=-L/2}^{L/2-1}J[1+\delta(-1)^x]\sum_{\alpha}(c^\dag_{x,\alpha}c^{}_{x+1,\alpha}+\hc),
\end{eqnarray}
where $c_{x,\alpha}^{(\dag)}$ are creation (annihilation) operators at site $x$, 
and  $n_{x,\alpha} = c_{x,\alpha}^{\dag}c^{}_{x,\alpha}$ represents the number operator.
We use Greek indices ($\alpha,\beta\in\{1,\dots, N\}$) to indicate the flavors.
The interaction term $H_U$ is parameterized by the on-site Hubbard coupling $U$, 
while the kinetic term $H_T$, is characterized by the hopping $J$, and the lattice dimerization parameter $\delta$.
For numerical calculations we keep the total number of sites $L$  even,
and use periodic boundary conditions (PBC).

We are interested in the dynamics of a dressed $N$-ion injected at $x=0$ in the lattice, in the limit of large Hubbard interactions $U/J\gg1$.
Its wave function resides mostly in the upper band, and has an energy $E_N\approx N(N-1)/2$.
In the absence of a dissipative mechanism, the $N$-ion has a long lifetime since it propagates through the lattice only through quantum fluctuations.

In order to construct the effective Hamiltonian that describes the 
dynamics of such an object in the limit of large $U$, the kinetic term $H_T$ is treated as a perturbation to $H_U$.
The effective Hamiltonian for its dynamics is obtained by projecting 
out the low-energy states of the rest of the $N$-particle states using $N$th order 
perturbation theory in $H_T$. Below we consider the particular SU(2) and SU(3) 
cases, and construct explicitly the leading-order effective Hamiltonians that describe the propagation of dressed doublons and trions, respectively, through the lattice.

\subsection{Effective Hamiltonian for doublons for the \texorpdfstring{SU(2)}{SU(2)} SSHH model}
\label{sec:SU2}
For the $SU(N=2)$ case in Eq.~\eqref{eq:SSHH}, the two flavors can be identified with the spin-$\uparrow$ and spin-$\downarrow$ projections for a spin-1/2, and the model reduces to the Hubbard model with a dimerized hopping.
The initial two-particle state is composed of two fermions with opposite spins, located at the same site,
\begin{equation}
|\Phi_x^{(2)}\rangle = \Phi_x^{(2)\dag}|0\rangle = c^\dag_{x\up} c^\dag_{x\down}|0\rangle
\equiv|x\!\up,x\!\down\rangle.\label{eq:Psi_doublon}
\end{equation}
Such a state has an energy $E_2\approx U$. 
To construct the effective model, we define the projector $P$ into the bare doublonic states (linear superposition of states of the form \eqref{eq:Psi_doublon}) and the projector $S$ 
into the rest of the biparticle scattering states, respectively,
\begin{equation}
P = \sum_{x}| \Phi_x^{(2)}\rangle \langle \Phi_x^{(2)}|,\quad 
S = \frac{1}{U}\sum_{x<y,\sigma}  
|x\sigma y\bar\sigma\rangle \langle x\sigma y\bar\sigma|,
\end{equation}
where $x,y$ run over lattice sites, and $\sigma,\bar\sigma$ are opposite spin projections. 
The effective Hamiltonian is obtained by projecting out the two-particle scattering states using a second order perturbation theory in $H_T$,
\begin{equation}\label{2nd_order}
H_{\rm eff} = PH_UP + PH_TP+ PH_TSH_TP+\dots.
\end{equation}
After some algebra, the effective Hamiltonian describing the dynamics is
constructed as 
\begin{eqnarray}\label{eq:HD}
H_{\rm eff} &=& \sum_{x=-L/2}^{L/2-1}
\bigg[U+\frac{4J^2(1+\delta^2)}{U}\bigg]\Phi_x^{(2)\dag} \Phi_x^{(2)} 
+ \frac{2J^2}{U}[1+\delta(-1)^x]^2(\Phi_x^{(2)\dag} \Phi_{x+1}^{(2)} +\hc)+\dots
.
\end{eqnarray}

\begin{figure}[t]
	\includegraphics[width=\textwidth]{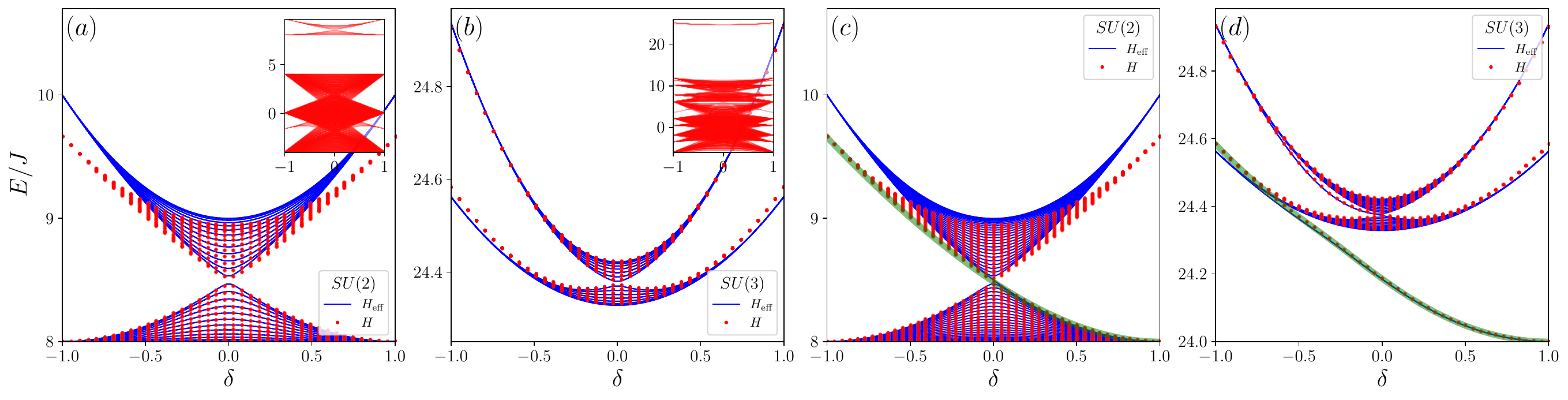}
	\caption{Comparison between the energy dispersion for the highest excited states of two interacting SSH models (red dots, $H$) and the dispersion obtained from their respective effective Hamiltonians (blue lines, $H_{\rm eff}$) as a function of the dimerization parameter for (a) and (b) periodic or (c) and (d) open boundary conditions.
	Panel (a) presents the SU(2) model's doublon bands in an $L=50$ chain, with an inset showing the entire many-body spectrum, while (b) presents the SU(3) model's trion bands in an $L=30$ chain, with the inset showing the entire many-body spectrum (for an $L=14$ chain).
	(c, d) For the same systems, but with open boundary conditions, both for the full Hamiltonian $H$ and the effective counterpart $H_{\rm eff}$, there are edge states (green thick line for $H$). In (c) the SU(2) model, the edge states stick to the bulk bands extrema and pass from the upper to lower doublon band at $\delta=0$, while in (d) the SU(3) model, they form below the trion bands for generic $\delta$. In all cases $U=8J$.}
	\label{fig:su2_3_eff}
\end{figure}

The model resembles the non-interacting SSH model, with a renormalized hopping and with additional on-site energy. 
The maximum propagation velocity of the two-particle state in the lattice follows readily, $v_2\simeq 4J^2(1-|\delta|)^2/U$.
In Fig.~\ref{fig:su2_3_eff}(a) we present a comparison for the doublonic band
energy spectrum obtained by diagonalizing the full Hamiltonian~\eqref{eq:SSHH}
and the one from the effective Hamiltonian Eq.~\eqref{eq:HD} 
for $U=8J$.
Apart from a slight deviation, that decreases with increasing U, the effective model provides a good estimate for the band structure. 

\subsection{Effective Hamiltonian for the trions for \texorpdfstring{SU(3)}{SU(3)} SSHH model}
\label{sec:su3}

Now we derive the  effective model that describes the dynamics of a dressed trion. 
Following the same logic of the previous section, 
the creation operator for the initial state at site $x$ reads 
$\Phi_x^{(3)\dag}=c_{x1}^\dag c_{x2}^\dag c_{x3}^\dag$ 
and a localized state $|\Phi_x^{(3)} \rangle = \Phi_x^{(3)\dag}|0\rangle$ has an energy $E_3\approx 3U$. 
The projector to the trion subspace is $P=\sum_{x}| \Phi_x^{(3)}\rangle \langle \Phi_x^{(3)}|$.

The rest of three-particle states correspond to configurations where each particle sits on different sites, and configurations where two particles form a dressed doublon while the other particle is scattered through the lattice.
Therefore, the projector $S$, orthogonal to $P$, is the sum of projectors $S_1$ and $S_2$ to the scattering and doublon states respectively,
\begin{equation}
S=S_1+S_2,\quad S_1 = \frac{1}{3U}\sum_{\substack{x<y<z\\ \alpha<\beta}}|x\alpha, y\beta, z\gamma\rangle
\langle x\alpha, y\beta, z\gamma|,\quad 
S_2 = \frac{1}{2U}\sum_{\substack{x\neq y\\\alpha<\beta}} |x\alpha,x\beta,y\gamma\rangle \langle x\alpha,x\beta,y\gamma|.
\end{equation}
The flavor index $\gamma$ is always the remaining third flavor, different from summed-over flavors $\alpha$ and $\beta$.

To construct the effective trionic Hamiltonian it requires to perform an expansion in the $3$rd order perturbation theory in $H_T$,
\begin{eqnarray}
H_{\rm eff} &=& PH_UP + P H_{T} P + PH_{T}SH_{T}P + PH_{T}SH_{T}SH_{T}P\notag\\ 
&&{}-\frac{1}{2}(PH_{T}PH_{T}S^2H_{T}P + PH_{T}S^2H_{T}PH_{T}P)+\dots.
\end{eqnarray}
The effective trionic Hamiltonian follows after a tedious, but otherwise straightforward calculation,
\begin{equation}\label{eff_trion}
H_{\rm eff} = \sum_{x=-L/2}^{L/2-1}\bigg[3U + \frac{3J^2}{U}(1+\delta^2)\bigg]\Phi_x^{(3)\dag} \Phi_{x}^{(3)}
+\frac{3 J^3}{2U^2}[1+\delta(-1)^x]^3 (\Phi_x^{(3)\dag} \Phi_{x+1}^{(3)}+\hc)+\dots,
\end{equation}
showing again that in this limit the trion Hamiltonian is mapped to a non-interacting SSH-type Hamiltonian, with an additional on-site energy. To this order, the parameters of effective SSH Hamiltonian from Eq.~(2) in the main text are
\begin{equation}
J_3=\frac32\frac{J^3}{U^2}(1+3\delta^2),\quad 
\delta_3 =\frac{3+\delta^2}{1+3\delta^2}\delta,\quad
\text{and }
E_3=3U+3\frac{J^2}{U}(1+\delta^2).
\end{equation}
Higher-order corrections,  will further renormalize the couplings such that the SSH Hamiltonian from Eq.~(2) in the main text is recovered. 
$H_{\rm eff}$ indicates that the maximum propagation velocity of the bound three-particle state in the lattice is $v_3\propto 3J^3(1-|\delta|)^3/U^2$.
Fig.~\ref{fig:su2_3_eff}(b) shows an almost perfect match between 
trion bands' energy dispersion obtained from Eq.~\eqref{eq:SSHH} and spectrum of the 
effective model~\eqref{eff_trion}. 

\subsection{Effective model for the dressed \texorpdfstring{$N$}{N}-ion in a lattice with \texorpdfstring{SU$(N)$}{SU(N)} symmetry}

The effective leading order Hamiltonian for the dressed $N$-ion is difficult to determine in closed form since it already requires an $N$th order perturbation theory in $H_T$.
Instead, we obtain an approximate leading order Hamiltonian from Eq.~\eqref{eq:SSHH} which captures first-order corrections to the kinetic and the on-site energy of $N$-ions.
The dominant contribution to the on-site energy is already obtained in 
the second order perturbation theory for the doublonic band.
In contrast, the dominant contribution to the hopping term requires 
the $N$th order in the perturbation theory, since it requires tunneling of 
all $N$ particles to an adjacent site.

Therefore, the approximate effective Hamiltonian reads
\begin{equation}
H_{\rm eff}\simeq\sum_{x=-L/2}^{L/2-1}
\bigg[\frac{UN(N-1)}{2}+ \frac{2N}{N-1}\frac{J^2(1+\delta^2)}{U}\bigg]\Phi_x^{(N)\dag} \Phi_x^{(N)}
+\frac{N J^N[1+\delta(-1)^x]^N}{(N-1)!U^{N-1}}(\Phi_x^{(N)\dag} \Phi_{x+1}^{(N)}+\hc)+\dots,
\end{equation}
where $\Phi_j^{(N)\dag} = c^\dag_{x1} c^\dag_{x2}\dots c^\dag_{xN}$ is the $N$-ion creation operator.
The neglected orders contributing to the on-site term in the perturbation theory behave as $1/U^{n-1}$ for $n<N$, and are generally larger than the hopping term correction.
Nevertheless, it is the latter which is responsible for the characteristic dimerization that allows to map the highest band in the SU($N$) model  the non-interacting SSH Hamiltonian.

\subsection{Edge states}
The effective models derived above are all formally equivalent to non-interacting SSH models with additional on-site energy which breaks the chiral symmetry.
However, the models present localized states at the edges due to reduced quantum 
fluctuations at the boundaries.
Such non-topological interaction-induced edge states have been seen also in two-body physics in bosonic SSH models~\cite{DiLiberto2016,Gorlach2017}.
Due to the additional variation of the effective chemical potential at the edge, topological invariants based on the Berry phase do not inform us on the edge physics in such systems, and zero-energy states are no longer pinned to zero energy. 
Nevertheless, the invariants remain good measures that characterize the bulk topology of the interacting Hamiltonian.

For open boundary conditions, the effective Hamiltonians are complemented 
by additional edge terms which take into account the renormalization of 
on-site energy at the edges. 
Then they reproduce well the exact diagonalization results of the many-body 
Hamiltonian~\eqref{eq:SSHH}.  The green lines in Fig.~\ref{fig:su2_3_eff}(c) 
and (d)] indicate the non-topological localized states at the edges, which for the SU(3) case 
decouple from the rest of the trionic band as the dimerization parameter 
$\delta$ is increased.

\section{Chiral symmetry, inversion symmetry, and the Berry phase quantization}
The chiral symmetry represented by operator $\Gamma$ acts on the fermion operators as follows~\cite{Chiu2016}:
\begin{equation}
\Gamma c_{x,\alpha} \Gamma^{-1} = (-1)^x c^\dag_{x,\alpha}, \quad \Gamma i \Gamma^{-1} = -i.
\end{equation}
Let us consider SU($N$) Hamiltonian~\eqref{eq:SSHH}, supplemented by a 
uniform chemical potential
\begin{equation}
H\to H-\mu\sum_{x,\alpha} n_{x,\alpha}.
\end{equation}
Under the chiral symmetry, the new Hamiltonian transforms as
\begin{eqnarray}
\Gamma H\Gamma^{-1} &=& \sum_x\bigg\{
\sum_{\alpha\neq\beta} \frac{U}{2}(1-n_{x,\alpha})(1-n_{x,\beta}) + \sum_\alpha 
\big[
J (1+\delta(-1)^x)(c^\dag_{x,\alpha} c^{}_{x+1,\alpha} + \hc)-\mu(1-n_{x,\alpha})
\big]
\bigg\},\notag\\
&=&\sum_x\bigg\{
\frac{U}{2}N(N-1)-\mu N+
\sum_{\alpha\neq\beta}\frac{U}{2}n_{x,\alpha}n_{x,\beta}\\
&&{}+\sum_{\alpha}
J (1+\delta(-1)^x)(c^\dag_{x,\alpha} c^{}_{x+1,\alpha}+\hc)+(\mu-(N-1)U)n_{x,\alpha}
\bigg\}\notag.
\end{eqnarray}
It follows that the interacting Hamiltonian is chiral symmetric $\Gamma H\Gamma^{-1}=H$ for a chemical potential tuned to
\begin{equation}
\mu = (N-1)U/2.
\end{equation}
The Hamiltonian is also invariant under the inversion symmetry. 
For open boundary conditions, the inversion symmetry acts as 
$Ic_{x,\alpha}I^{-1}=c_{-x-1,\alpha}$, with inversion symmetry center taken 
as the middle position between sites $x=-1$ and $x=0$.
For a finite $U$, the inversion symmetry is preserved, while the chiral symmetry is generally broken.
Nevertheless, inversion symmetry is sufficient for the many-body Berry phase to be quantized.
Under spatial inversion, $\gamma_B$ transforms as
\begin{eqnarray}
\gamma_B\to -\text{Im}\log \prod_n \det[S^{(n+1,n)}(-X)] = -\text{Im}\log \prod_n \det[S^{(n,n+1)}(X)]^* = -\gamma_B\, (\text{mod } 2\pi).
\end{eqnarray}
Note the interchange of $n$ and $n+1$ in the first step, compared to the original formula. This is due to inversion changing $k_n\to -k_n$, where $k_n$ is the momentum associated to $\theta_n$, $k_n=2\pi n/ML$, for $n\in\{0,\dots,L-1\}$.
Then a shift by $2\pi/L$, brings the momenta back in the first Brillouin zone $[0,2\pi/L)$, but expectation values are taken in opposite order.
In the present case, the system has inversion symmetry, therefore $\gamma_B = -\gamma_B\,(\text{mod } 2\pi)$. This is possible only if $\gamma_B=0$ or $\pi$ (mod $2\pi$). This indicates that the Berry phase is quantized and there are two possible topological phases. 
We have shown that, similar to the non-interacting model, the interacting model 
exhibits both phases, and a topological transition exists between them at $\delta=0$.

The robustness of $\gamma_B$ was checked against moderate on-site and hopping disorder that break the inversion symmetry.
Hopping disorder or on-site disorder, respectively, are introduced in Eq.~\eqref{eq:SSHH} as
\begin{equation}
H \to H + \sum_{x=-L/2}^{L/2-1}\delta J_{x,x+1}(c^\dag_{x,\alpha}c^{}_{x+1,\alpha}+\hc),
\text{ or } H \to H + \sum_{x=-L/2}^{L/2-1}\delta\mu_{x}^{}c^\dag_{x,\alpha}c^{}_{x,\alpha},
\end{equation}
with $\delta J_{x,x+1}\in W[-0.5,0.5]$, and $\delta\mu_x\in W[-0.5,0.5]$, randomly chosen from a uniform distribution with $W$, the disorder amplitude. The disorder-averaged value of the Berry phase $\avg{\gamma_B}$ is usually shown in these cases.

\section{Mean chiral displacement}
\begin{figure}[t]
	\includegraphics[width=0.65\columnwidth]{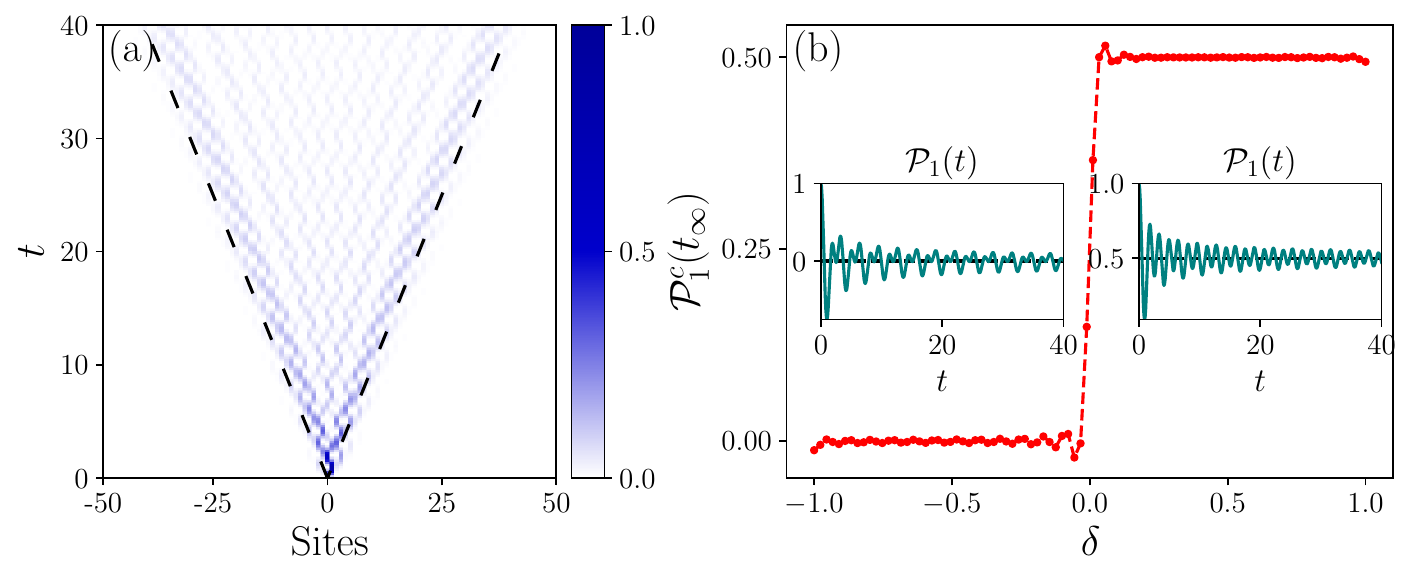}
	\caption{ Single-particle quantum walk in the non-interacting spinless SSH model. 
	(a) The time evolution of the density profile $\langle n(x,t)\rangle$ at $\delta=0.5$. 
	The dashed black lines denote the wave front of the propagation with a velocity
	$v_1=2J(1-|\delta|)$. 
	(b) Cumulative MCD $\mathcal P^c_1(t_\infty)$ as a function of the dimerization parameter $\delta$. (Insets) Characteristic time evolution of $\mathcal{P}_1(t)$ for $\delta=-0.5$ (left) and $\delta=0.5$ (right).
	In the trivial (topological) regime $\mathcal P_1(t)$ converges asymptotically to $0\, (0.5)$. In simulations, we consider $100$ sites lattices and the particle is initially injected in the middle of the chain. 
	Time is measured in units of $1/J$.}
	\label{fig:noint}
\end{figure}
Here we discuss in more detail the single particle and the many-body MCD. 
First, we present results for the non-interacting SSH model, and then discuss  the doublonic and trionic many-body MCD.

When a single particle is injected in the lattice, the interactions play no role and the underlying lattice Hamiltonian reduces simply to the non-interacting spinless SSH model~\cite{Su1979}. 
The lattice is bipartite, each unit cell contains two sites, labeled $A$ and $B$,
and the chiral operator is $\Gamma=\sigma^z$ within the site basis. 

Following the quench, the system state at $t=0$ becomes $| \Psi(t=0)\rangle = c^\dagger_{0,\beta} \, |0\rangle$, with $\beta$ a sublattice index. 
The wave function follows a unitary evolution in time according to the Schr\"odinger equation 
$| \Psi(t)\rangle =\exp(-iHt) |\Psi(t=0)\rangle$. 
The density profile $\langle n(x,t)\rangle$ develops a light-cone propagation, with a maximum
Lieb-Robinson velocity~$v_1=2J(1-|\delta|)$.
A typical behavior for $\langle n(x,t)\rangle$ in the topological regime ($\delta>0$) is displayed in Fig.~\ref{fig:noint} (a). This behavior remains qualitatively the same in the trivial regime, as $\langle n(x,t)\rangle$ does not differentiate between the two phases. 
The quantity that discriminates between the two regimes is the MCD.
In the present setup, we evaluate it as 
\begin{equation}
	\label{eq:P1_0}
	{\cal P}_1(t) = {\sum_x}'x
	\sum_{\alpha, \beta, \gamma}^{\{A,B\}} \langle 0|c_{0,\alpha} e^{i Ht} 
	c^\dagger_{x, \beta} \sigma^z_{\beta\gamma} c_{x, \gamma} e^{-iHt}c^\dagger_{0,\alpha}|0\rangle,
\end{equation}
where the primed sum over $x$ indicates that the sum is performed over the lattice unit cells, 
and the second sum is over the sublattice labels. 

\begin{figure}
	\includegraphics[width=\textwidth]{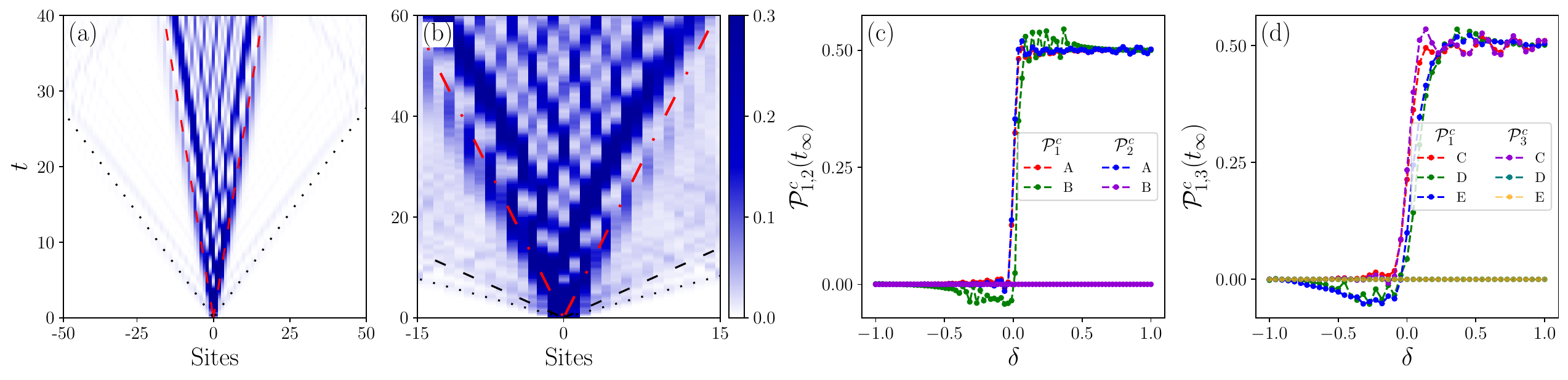}
	\caption{Time evolution of the density profile in a multiparticle quantum walk in (a) the SU(2), and (b) SU(3) models. 
	Dotted lines denote propagation front corresponding to a free particle maximum velocity $v_1$ in the lattice, dashed lines, the dressed doublon state's maximum velocity $v_2$, and dash-dot lines, the dressed trion state's maximum velocity $v_3$, estimated in the effective models.
	In time, we see in (a) and (b) that a fraction of the dressed $N$-ion state density breaks away and propagates with faster velocity $v_n$, $n<N$, corresponding its possible subsystems.
	The simulation parameters are (a) $L=100$, $U=8J$, $\delta=0.1$, and (b) $L=30$, $\delta=0.1$ $U=3J$. The density recorded is artificially capped at $0.3$, such that the breakaway probability amplitudes with $v_n$ are visible. Time is in units $1/J$.
	The cumulative average MCD for several initial conditions (A, B, C, D, E) for (c) SU(2), and (d) SU(3) models, for $L=100$, $U=8J$, $t_\infty=40$, and $L=40$, $U=3J$, $t_\infty=20$, respectively. (c) In the SU(2) model, setup A consists in injecting a two particle at $x=0$, and B, a pair of spin up and down at different sites near $x=0$. The panel shows $\mathcal P_1^c(t_\infty)$ and $\mathcal P_2^c(t_\infty)$ for A and B. 
	(d) In the SU(3) model, setup C consists in injecting three particles at $x=0$, D, three particles at three different adjacent sites around $x=0$, and E, two particles at $x=0$ and a single particle at a different site near $x=0$.
	The panel shows $\mathcal P_1^c(t_\infty)$ and $\mathcal P_3^c(t_\infty)$ for C, D, and E. 
	While $\mathcal P_1^c$ tends close to 0 or 0.5 in all setups, $P_2^c$, and $\mathcal P_3^c$ exist only when two, and three particles are injected in the lattice at the same site, respectively. 
	Errors in $P_1^c$ appear at small $|\delta|$ due to single particle excitations reaching the lattice edges.
	}
	\label{fig:ldos_su2_3}
\end{figure}

The numerical results are confirmed by evaluating the MCD in Eq.~\eqref{eq:P1_0} analytically.
Since the model is non-interacting, we compute the average in \eqref{eq:P1_0}, by using the Wick's theorem,
in terms of the non-interacting Green's functions. 
Introducing the greater Green's functions
$iG_{\alpha\beta}^{> (0)}(t) = \langle c_{x, \alpha} (t) c^\dag_{0, \beta}(0)\rangle $, the MCD reduces 
to  $ {\cal P}_1(t) = \sum_x ' x \mathrm{Tr}\big \{ G^{> (0)} (-x, -t) \sigma^z G^{>(0)}(x,t) \big \}$. 
By performing the Fourier transform we get 
\begin{equation}
	{\cal P}_1(t) = i \int {dk\frac 2\pi} \text{Tr}\big \{ G^{> (0)} (k, -t) \sigma^z 
	\partial _k G^{>(0)}(k,t) \big \}.
\end{equation}
A straightforward calculation shows that the MCD is
\begin{equation}
{\cal P}_1(t)\simeq \frac\nu 2 -\int {dk\frac 4\pi} \cos(2 E_k t) (\bm n_k \times \partial_k \bm n_k)_z
,
\end{equation}
with the winding number $\nu=\int \frac{dk}{2\pi} (\bm n_k\times \pd_k\bm n_k )_z$, $\bm n_k=\bm h_k/|\bm h_k|$, and $H=\bm h_k\cdot\bm\sigma$.
In the long time limit ${\cal P}_1(t)\to \nu/2$, as the highly oscillating correction
averages to zero for $t_\infty\to \infty$, thus confirming our numerical results. 
The Zak phase $\gamma_Z=\pi\nu$ (modulo $2\pi$), so it will not be able to distinguish nontrivial phases characterized by an even $\nu$.

For a single-particle quantum walk the theoretical findings corroborate the experimental measurements for ${\cal P}_1(t)$ performed in photonic lattices~\cite{Cardano2017,Wang2019} 
thus confirming that ${\cal P}_1(t)$ is able to capture the bulk states' topology (see Fig.~\ref{fig:noint}). 
Evaluating analytically ${\cal P}_1(t)$ when a doublon or a trion are injected into the lattice is not an easy task, since it requires evaluation of various combinations of full Green's functions.
The presence of strong on-site interaction requires a careful analysis, and there is no guarantee that perturbation theory works for large $U$.

In SU($N$) models, a fraction of a dressed $N$-ion amplitude can break away due to quantum fluctuations. 
These excitations will propagate at a faster maximum velocity due to a smaller mass.
Figure~\ref{fig:ldos_su2_3} shows the time evolution for the density profile for (a) SU(2) and (b) SU(3) models.
We identify, for the SU(2) case, excitations propagating with the single-particle velocity $v_1$, while for an SU(3) model, there are offshoots with velocities corresponding to single particles, $v_1$, and to dressed doublons, $v_2$.
The figures eliminate all amplitudes above a certain threshold, such that the small losses propagating with $v_1$ or $v_2$ are visible. In the SU(2) model we record $\mathcal P_1(t)$ and ${\cal P}_2(t)$, defined as 
\begin{equation}
{\cal P}_2(t)
=
\langle n_2(t) \rangle^{-1}
{\sum_x}'
\langle \Psi(t) |
\Phi^{(2)\dagger}_{x} (-1)^xx \Phi^{(2)}_{x}
| \Psi(t) \rangle,
\end{equation}
while in the SU(3) model, $\mathcal P_1(t)$ and $\mathcal P_3(t)$, in Fig.~\ref{fig:ldos_su2_3}(c) and (d), respectively.
In simulations, we also inject particles in the lower energy bands. This is done in SU(2) model by injecting a pair of particles at different sites. Then, modulo Hubbard local repulsion between them, the two particles propagate freely, such that $\mathcal P_1(t)$, properly normalized, has a response similar to the noninteracting SSH model.
Similarly, in the SU(3) model, we inject the particles at  three different sites, or as a doublon plus a single particle. In all cases we see that $\mathcal P_1(t)$ approximates the quantized response in the noninteracting SSH model.
In these cases, due to repulsive Hubbard interactions, formations of a doublon in SU(2) model or a trion in SU(3) is suppressed such that (unnormalized) $\mathcal P_2(t)$ and $\mathcal P_3(t)$, respectively, are zero [datasets B, D, E in Fig.~\ref{fig:ldos_su2_3}(c) and (d)].
Therefore, $\mathcal P_2(t)$ and $\mathcal P_3(t)$ are specific measures when injecting a localized two-particle state in the SU(2) model, or a three-particle one in SU(3) model.

\begin{figure}[t]
\includegraphics[width=\textwidth]{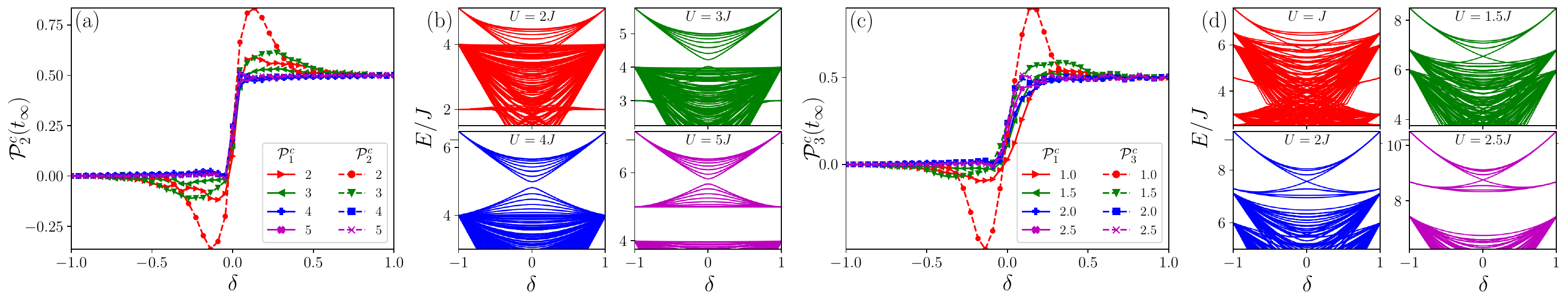}
\caption{(a) The cumulative average MCD $\mathcal P_1^c(t_\infty)$ and $\mathcal P_2^c(t_\infty)$ at four interaction amplitudes for the SU(2) SSH model. A doublon is injected at the lattice center. The simulation time is $t_\infty=40/J$, the lattice size is $L=100$, and the legend indicates the interaction strength $U/J$. (b) Part of the many-body spectrum comprising the doublon band as a function of the dimerization parameter $\delta$, for $L=30$.
(c) $\mathcal P_1^c(t_\infty)$ and $\mathcal P_3^c(t_\infty)$ for  the SU(3) SSH model for a system size $L=40$. (d) The corresponding upper part of the many-body spectrum containing the trion band, as a function of $\delta$. Lattice size is fixed to $L=12$.
In (c) a trion is injected at the lattice center. The simulation time is $t_\infty=12/J$ for $U=J$ and $1.5J$, $t_\infty=20/J$ for $U=2J$ and $2.5J$, and the legend indicates the interaction strength $U/J$.
The MCD is quantized when the doublon and trion bands are energetically separated from the rest of the many-body spectrum in SU(2) and SU(3) models, respectively.
}
\label{fig:diff_U}
\end{figure}

Notably, MCD appears to be well quantized under weak interactions. This is demonstrated in Figure~\ref{fig:diff_U}, where a doublon and a trion are injected at the center of the lattice for both SU(2) and SU(3) models. The cumulative average MCDs $\mathcal P^c_{1,2}(t_\infty)$ and $\mathcal P^c_{1,3}(t_\infty)$ are recorded for various interaction strengths $U$. The results indicate that MCD is strongly affected by the quantum fluctuations for certain $\delta$ values, namely when the doublon or trion band overlaps in energy with the rest of the many-body spectrum in the SU(2) or SU(3) model, respectively. In all other cases, the MCD shows the expected transition from 0 in the trivial region to 0.5 in the nontrivial region, even when there is no energy gap, as shown in Figs.~\ref{fig:diff_U}(a) and~\ref{fig:diff_U}(b).

\section{Two trions}
\begin{figure}[t]
\includegraphics[width=0.65\textwidth]{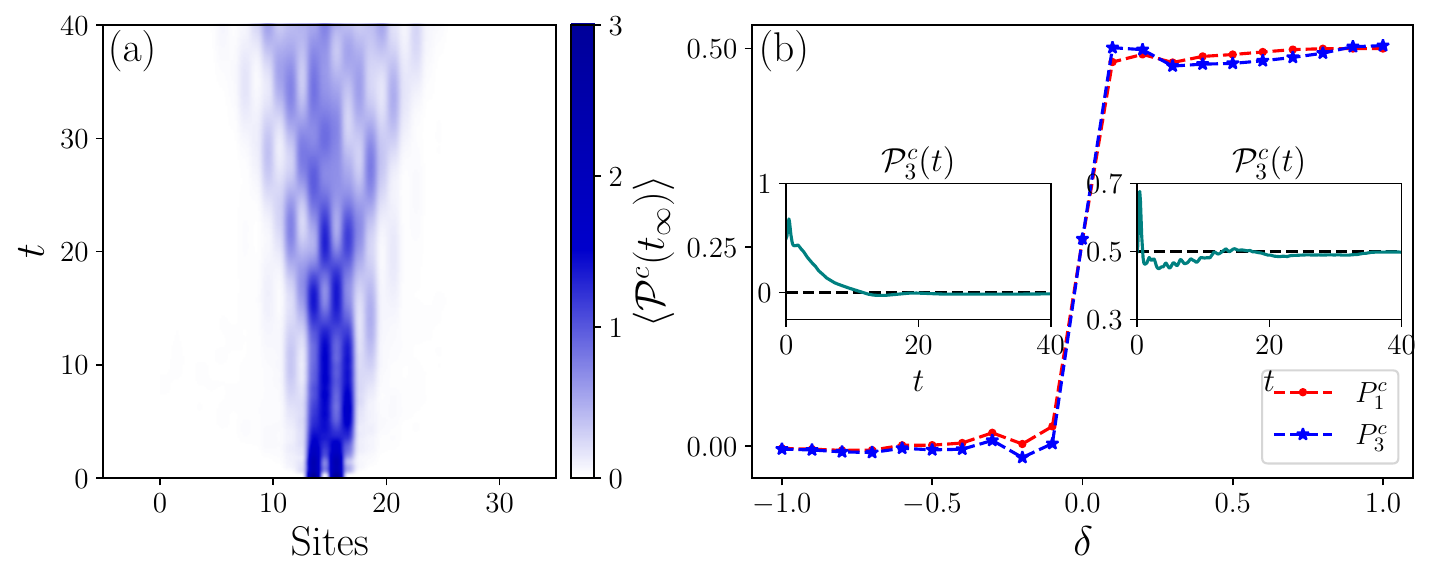}
\caption{(a) Time-evolution of the density profile for two trions at $\delta=0.2$.
(b) Cumulative average MCD $\mathcal P_1^c(t_\infty)$ and $\mathcal P_3^c(t_\infty)$ for two dressed trions.
Left and right insets represent the full time evolution of $P_3^c(t)$ at $\delta=-0.2$ and $\delta=0.2$, respectively.
The two trions are injected at site 14 and 16 in an $L=30$ lattice. 
[$U=3J$ and time $t$ is in units of $1/J$.]}
\label{fig:2trions}
\end{figure}

We have also addressed the case when two three-particle states are injected in the lattice. 
Numerical results for the average density $\langle n(x,t)\rangle$ and for the chiral and cumulative MCD $\mathcal P_{1,3}(t)$, using TEBD, are presented in Fig.~\ref{fig:2trions} for a disorder-free lattice. 
The results show that the cumulative MCD remains a good indicator of topological properties.
Still, because the lattice is populated by more than a single trion, there is an effective interaction of the form 
\begin{equation}
H_{\rm eff}^{\rm int}\approx -|V_3| \sum_{\langle x,x'\rangle} \Phi_x^{(3)\dag} \Phi_x^{(3)} \Phi_{x'}^{(3)\dag} \Phi_{x'}^{(3)} = -|V_3|\sum_{\langle x,x'\rangle } n_x^{\Phi^{(3)}}n_{x'}^{\Phi^{(3)}},
\end{equation}
with $n_x^{\Phi^{(3)}}=  \Phi_x^{(3)\dag} \Phi_x^{(3)}$ the trion number operator, and $V_3/J_3\sim U/J$. 
Therefore, in the strong interaction limit, the three-particle states introduced at different positions in the lattice become heavy and  attract each other.
Furthermore, higher-order processes generate a residual interaction between the dressed trions and doublons as well, producing a diffusion of dressed doublons into the lattice~\cite{werner2022spectroscopic}.

\section{Many-body Berry phase}
\begin{figure}[th]
	\includegraphics[width=0.75\textwidth]{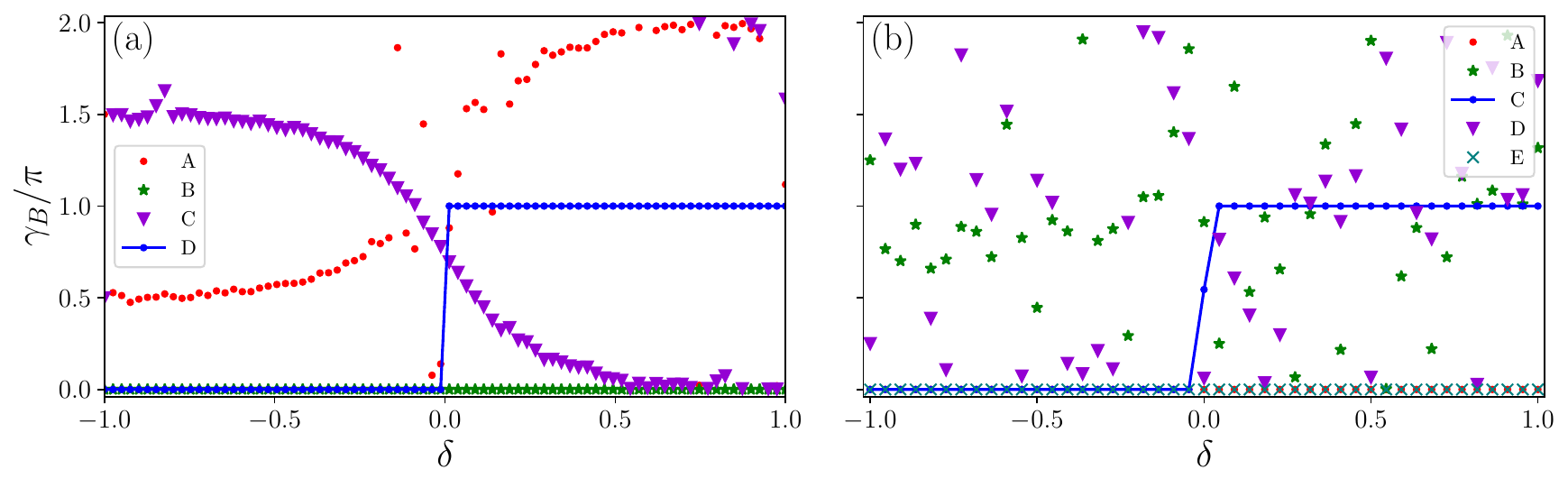}
	\caption{Many-body Berry phases in the (a) interacting SU(2) and (b) SU(3) models when summing Berry phases for various ranges of many-body states starting from the ground state for clean $L=8$ chains and $U=8J$. 
		In panel (a), B (green) represents integration over all low-energy states, including the lower doublon band, showing that a $\pi$-jump is not detectable in $SU(2N)$ models. A (red) and C (violet) are the results when subtracting the highest state or adding a new highest state in energy to the integration in B. A $\pi$-jump is obtained in D (blue), when numerically imposing twisted boundary conditions only for spin up. 
		In panel (b), we show the constant $\gamma_B$ obtained when summing over all non-trionic states, A (red), or over all three-particle states, E (teal). Summing all states including the lower trion band, C (blue), shows the expected transition at $\delta=0$. Finally, subtracting or adding a single state to C, in B and D, respectively, gives an unquantized $\gamma_B$.}
	\label{fig:bp_sm}
\end{figure}

The many-body Berry phase was defined in Eqs.~(3)~and (4) in the main text.
It allows investigating numerically the topological properties of the interacting models in equilibrium.
The Berry phase shows however limitations in describing the topological properties for SU($N$) Hamiltonians with an even number of flavors $N=2n$.
Such a case is exemplified in Fig.~\ref{fig:bp_sm} for SU(2), with two interacting fermions of opposite spin projection.
The problem is similar to the noninteracting case, where the spinful model is just two decoupled chains for each spin projection.
Since there is $\pi$-jump at $\delta=0$ for each chain, there is $2\pi$ jump in the total model.
Since the Berry phases are defined modulo $2\pi$, it is impossible to distinguish in the Berry phase between $\delta<0$ and $\delta>0$ regimes.
This is generic for $SU(2n)$ noninteracting systems, since the MCD is proportional to the winding number $\nu$, while the Berry phases are defined $\pi\nu$ (modulo $2\pi$), and cannot discriminate phases with even $\nu$ from the trivial state. 
This remains true in the interacting SU(2) models (see Fig.~\ref{fig:bp_sm}(a), dataset B), where the Berry phase is zero at all $\delta$, while $\mathcal P_2(t)$ correctly distinguishes in Fig.~\ref{fig:ldos_su2_3}(c) the topological phases. 

Nevertheless, it is still possible to reveal the Berry phase $\pi$-jump at the transition for interacting models. Adding twisted boundary conditions for only one spin channel, allows one to observe the $\pi$-jump in the model even if interactions mix the two flavors (see Fig.~\ref{fig:bp_sm}, dataset D).
We conclude that without such numerical probes, it will not be possible to record the topological transition in the Berry phase for $SU(2n)$ models.

The case of $SU(2n+1)$ is expected to show an unambiguous $\pi$-jump in $\gamma_B$ at $\delta=0$. Here we add further proof to the results in the main text for the SU(3) model (see Fig.~\ref{fig:bp_sm}(b)).
We show that the integration of the non-trion, low-energy states, gives a constant Berry phase $0$, Fig.~\ref{fig:bp_sm}(b), dataset A.
Therefore carrying integration for all states from ground state to the dimerization gap (including the lower trion band) shows the expected $\pi$-jump, seen also in the main text from integration only over the lower trion band.
We also show in Fig~\ref{fig:bp_sm}(b) that adding or subtracting even a single state to the lower trionic band results in an unquantized Berry phase with no trace of the jump. 
This is a more general result, which holds for most subsets of states, and shows that the many-body Berry phase has the expected quantization characteristic for an SSH model, only when integrating over the states visited by the dressed $N$-ion.

In the limit of large $M$ the determinant formula from Eq.~(3) in the main text converges to a sum of Berry phases over the many-body states,
\begin{eqnarray}
\gamma_B=-\text{Im}\ln\prod_{n=0}^{M-1}\det[S^{(n,n+1)}] = -\text{Im}\sum_{n=0}^{M-1} \tr \ln[S^{(n,n+1)}]
\simeq -\text{Im}\sum_{n=0}^{M-1} \sum_j \ln\avg{\Psi^{(n)}_j|e^{2\pi iX/ML}| \Psi^{(n+1)}_j},
\end{eqnarray}
where $j$ indexes states in the chosen subset of many-body states.
The last approximate expression is obtained in the limit of large $M$, in which the exponential operator is close unity, and states $\psi^{(n)}_j$ and $\psi^{(n+1)}_{j'}$ are nearly orthogonal for $j\neq j'$, such that the overlap matrix $S$ is close to a diagonal matrix.
Therefore, in the limit of large $M$, it follows that
\begin{equation}
\gamma_B \simeq -\sum_j \text{Im}\ln \prod_{n=0}^{M-1}\avg{\Psi^{(n)}_j|e^{2\pi i X/ML}| \Psi^{(n+1)}_j}.
\end{equation}
Equality is used everywhere modulo $2\pi$.
The latter $\gamma_B$ corresponds thus to a sum of Berry phases for the excited states of $M$ trions in an $ML$ lattice, where the trions in different lattices of size $L$ do not interact with each other~\cite{Resta1999, Resta2000}.

We find numerically that the determinant formula is more robust, and for disorder-free systems the single-point Berry phase $M=1$ is sufficient to identify the topological phases, and shows good quantization.
In the interacting disordered models, exact diagonalization is restricted to small sizes $L$. 
In such cases, increasing $M$ is useful to suppress finite size effects, as the effective length of the system is $ML$.

\bibliographystyle{apsrev4-2}
\bibliography{bibl}

\let\addcontentsline\oldaddcontentsline